\documentclass{vgtc}                          

\graphicspath{{figures/}{pictures/}{images/}{./}} 

\usepackage{times}                     

\usepackage{tabu}                      
\usepackage{booktabs}                  
\usepackage{lipsum}                    
\usepackage{mwe}                       
\usepackage{enumitem}

\usepackage{mathptmx}                  

\onlineid{0}

\vgtccategory{Research}

\vgtcinsertpkg

\title{Evaluating Closed-Loop EEG Feedback for Simulated Prosthetic Vision in Immersive VR: A Sham-Controlled Feasibility Study}

\author{Ruyi Cao\\ %
        \scriptsize UC Santa Barbara %
\and Lily M. Turkstra\\ %
        \scriptsize UC Santa Barbara %
\and Adyah Rastogi\\ %
     \scriptsize UC Santa Barbara %
\and Michael Beyeler\thanks{Correspondence: mbeyeler@ucsb.edu}\\ %
     \parbox{1.4in}{\scriptsize UC Santa Barbara}}

\teaser{
  \centering
  \includegraphics[width=.95\linewidth]{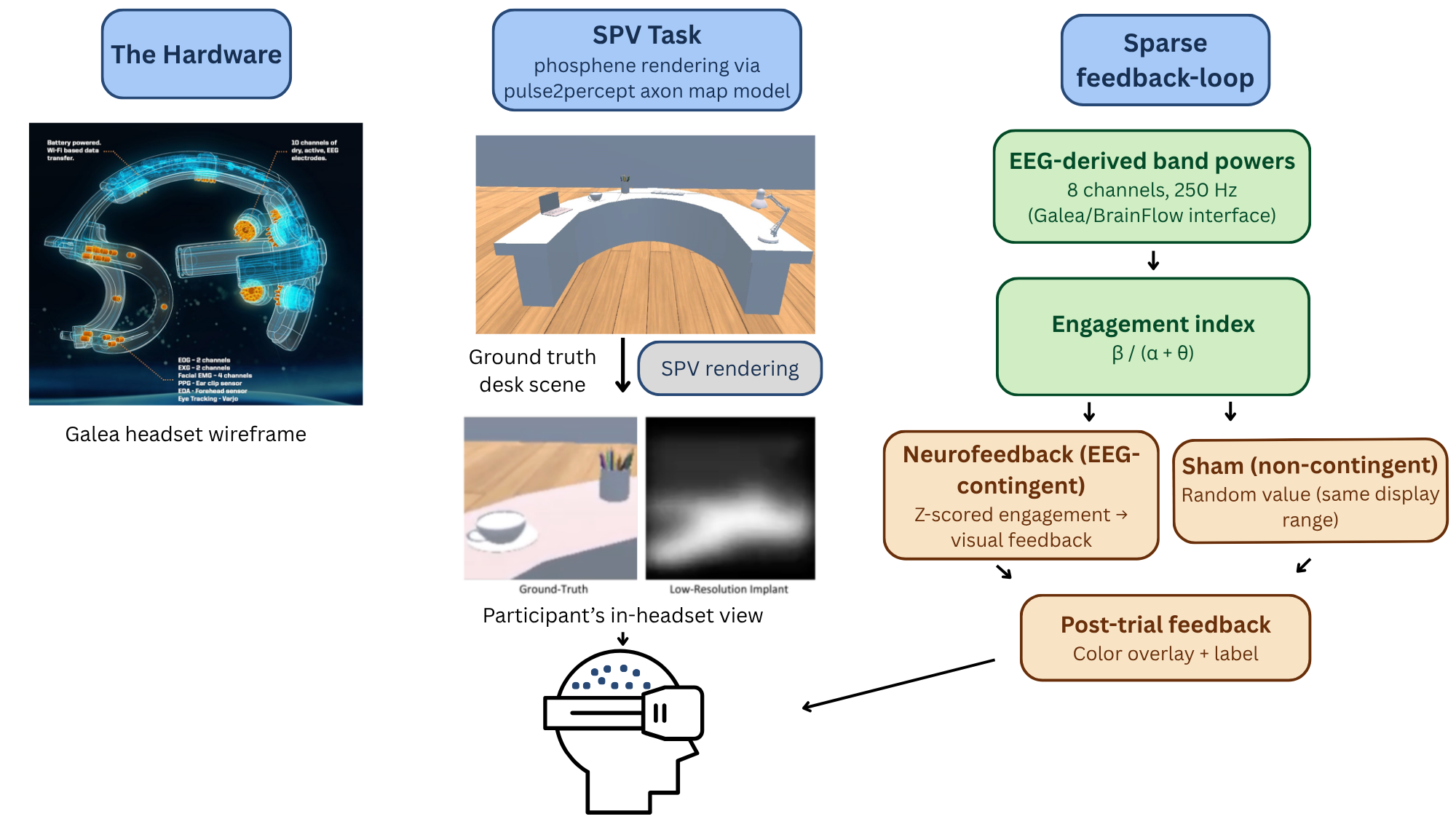}
  \caption{Closed-loop EEG-guided training platform for simulated prosthetic vision (SPV). 
    The Galea system combines an eight-channel dry-electrode EEG sampled at 250\,Hz with a head-mounted VR display. A panoramic virtual desk scene is rendered in real time as a low-resolution ($10\times10$) phosphene percept deployed in Unity. During each trial, the participant searches the SPV scene and selects the perceived target location. EEG-derived theta-, alpha-, and beta-band power estimates are combined into an engagement index, $\beta/(\alpha+\theta)$, and standardized relative to the participant's pre-training baseline. After each training trial, participants receive either feedback contingent on their own engagement score or visually matched non-contingent sham feedback. Feedback is presented only between trials to avoid competing with the degraded visual display during search.}
  \label{fig:teaser}
}

\abstract{
    Visual prostheses require users to interpret sparse and distorted artificial percepts through active visual search. We developed an EEG-guided neuroadaptive training platform for simulated prosthetic vision in immersive virtual reality and evaluated its feasibility in a sham-controlled object-localization task. Twenty-two sighted participants searched a virtual desk scene rendered through a low-resolution phosphene simulation while EEG was recorded using a dry-electrode headset integrated with a head-mounted display. During training, participants received post-trial visual feedback based either on a commonly used EEG engagement index, $\beta/(\alpha+\theta)$, or on visually matched non-contingent sham values. Both groups showed comparable within-session improvements in localization performance, consistent with practice, increasing familiarity with the simulated percepts, or refinement of search strategies. EEG-contingent feedback did not produce reliable group-level benefits in localization accuracy, completion time, workload, or modulation of the targeted index. Exploratory analyses showed substantial individual variability, but did not establish a feedback-specific relationship between the engagement index and behavioral performance. These findings demonstrate the feasibility of integrating EEG-contingent feedback with immersive simulated prosthetic vision, while identifying important limitations of the EEG measure, single-session training protocol, and post-trial feedback design.
} 


\begin{document}


\firstsection{Introduction}

\maketitle

Visual neuroprostheses, or \emph{bionic eyes}, aim to restore a rudimentary form of sight by translating visual input into patterns of electrical stimulation in the retina, optic nerve, or visual cortex~\cite{weiland_electrical_2016,fernandez_development_2018}. Clinically deployed systems can support basic navigation, object localization, and visually guided behavior~\cite{luo_argusr_2016,holzSubretinalPhotovoltaicImplant2025,karapanos_functional_2021}, but the resulting percepts remain sparse, low-resolution, and spatially distorted. Prosthetic vision is therefore not simply a pixelated version of natural sight. Users must actively scan the environment, interpret ambiguous phosphene patterns, maintain spatial orientation, and decide when a percept is reliable enough to guide action. The bottleneck is not only visual resolution; it is also the sustained cognitive effort required to make degraded artificial vision useful.

Simulated prosthetic vision (SPV) in immersive virtual reality offers a practical way to study this bottleneck before testing new strategies with small and heterogeneous implant cohorts. In SPV, sighted participants view virtual environments through phosphene-based renderings that approximate selected constraints of current and near-future visual prostheses~\cite{hayes_visually_2003,dagnelie_real_2007,kasowski_immersive_2022,thorn_virtual_2021,kasowski_simulated_2025}. SPV cannot reproduce the lived experience of blindness, long-term adaptation to an implant, or the idiosyncrasies of electrical stimulation. It can, however, support controlled and repeatable experiments that would be difficult to isolate clinically. Immersive VR is particularly useful because the degraded display is embedded in an active sensorimotor task: users must move their head, search the scene, and localize objects under the constraints imposed by the simulation.

Most SPV work has focused on improving the information delivered to the user. Image-processing and scene-simplification methods can emphasize edges, depth, semantic objects, or task-relevant structure~\cite{dagnelie_real_2007,vergnieux_simplification_2017,sanchez_garcia_semantic_2020,han_deep_2021,mccarthy_mobility_2014,sadeghiBenefitsThermalDistancefiltered2024,rasla_relative_2022}. These approaches treat prosthetic vision primarily as an encoding problem: because the display can render only a small fraction of the scene, the system should choose better information to show. But even with better encoding, users must learn how to search, attend, and interpret unfamiliar artificial percepts. Therefore, a complementary question is whether the training system can adapt to the user's cognitive state rather than treating the user as a passive endpoint of the display pipeline.

Neuroadaptive XR provides a framework for testing whether physiological signals can inform training during demanding immersive tasks. EEG has been used to derive measures associated with workload, attention, and task engagement~\cite{tremmel2019,berka2007,zahabi2025}. One commonly used composite measure is the engagement index, defined as beta power divided by the sum of alpha and theta power, $\beta/(\alpha+\theta)$~\cite{pope1995,berka2007}. In this study, we use the term \emph{engagement index} to refer specifically to this operational EEG measure. We do not assume that it uniquely or comprehensively measures attention, motivation, workload, or a unitary cognitive state. Rather, we test whether feedback based on this index can be implemented during SPV training, whether the index changes under contingent feedback, and whether it covaries with behavioral performance. These questions are particularly uncertain during active VR use, where dry-electrode EEG may be affected by head movement, muscle activity, and changes in electrode contact~\cite{rogala2016,tipple2024}.

We therefore developed a closed-loop platform combining a dry-electrode EEG headset, a head-mounted VR display, and a Unity-based SPV object-localization task (Fig.~\ref{fig:teaser}). Participants searched a panoramic desk scene rendered through an axon-map phosphene model. During training, the neurofeedback group received a color-coded post-trial display based on its own EEG-derived engagement index, whereas the sham group received visually matched non-contingent feedback. Feedback was sparse rather than continuous so that it would not introduce an additional visual signal during the already demanding SPV search.

This feasibility study asks three questions. First, how does localization performance change over a single session using the integrated EEG--SPV platform? Second, does EEG-contingent feedback produce greater behavioral improvement or modulation of the targeted index than visually matched sham feedback? Third, does the engagement index show exploratory associations with localization performance or substantial variation across participants?

Our contributions are: 
\begin{enumerate}[itemsep=0pt,topsep=0pt,parsep=0pt]
    \item an integrated closed-loop NeuroXR platform for studying cognitive-state-aware SPV training;
    \item a sham-controlled evaluation of sparse EEG-contingent feedback; and
    \item an analysis that separates the group-level intervention results from explicitly exploratory brain--behavior associations and individual trajectories.
\end{enumerate}

\section{Methods}

\subsection{Participants and Design}

Twenty-two sighted adults (15 female, 7 male; age: M = 20.55, SD = 1.77, range = 19-27) were recruited from the university community, and randomly assigned to either the neurofeedback (n = 12) or sham (n = 10) group. One participant was excluded before analysis because a computer failure during the pre-training phase corrupted the baseline EEG distribution used to generate feedback. The final sample comprised 22 participants, with 12 assigned to neurofeedback and 10 to sham. Three additional participants who completed earlier pilot versions of the protocol were not included in the formal sample. All participants reported normal or corrected-to-normal vision and no history of neurological or psychiatric disorders. The study was approved by the institutional review board, and all participants provided written informed consent.

Each participant completed a single session with three phases: 20 pre-training trials without feedback, 90 training trials divided into six blocks of 15, and 20 post-training trials without feedback. During training, the neurofeedback group received EEG-contingent feedback and the sham group received non-contingent feedback with the same timing and visual presentation. A short break was provided near the midpoint of training to reduce fatigue and discomfort from the combined EEG--VR headset.

The study was designed as an initial feasibility evaluation, and no a priori power analysis was conducted because no established effect-size estimate was available for EEG-contingent feedback during SPV training. The sample was therefore not intended to provide a definitive test of efficacy or to rule out small or moderate intervention effects.

\subsection{Apparatus and EEG Acquisition}

The experiment was implemented in Unity and presented through a Varjo XR-3 head-mounted display integrated with a Galea dry-electrode EEG headset. EEG was recorded from eight channels (F1, F2, C3, C4, P3, P4, Cz, and Pz) at 250~Hz. The experimental software accessed EEG-derived features through the Galea/BrainFlow interface and obtained online estimates of theta- (4--8~Hz), alpha- (8--13~Hz), and beta-band (13--30~Hz) power. The lower-level implementation of the online feature extraction, including filtering, spectral-estimation settings, and automated artifact handling, was not exposed to the experiment software. For each trial, the available band-power estimates were averaged across the eight channels and across the duration of the trial.
The Galea/BrainFlow interface provided online estimates of theta-, alpha-, and beta-band power, but did not expose the underlying preprocessing pipeline to the experiment software.

Averaging across all eight channels produced a single low-dimensional control signal suitable for online feedback, but removed spatial information that might distinguish regional neural activity from broadly distributed physiological or movement-related signals. Because trials were self-paced, the estimates were also averaged over recordings of different durations. Longer trials may therefore have produced more stable estimates than shorter trials.

\subsection{Simulated Prosthetic Vision}

Visual input was rendered using a psychophysically validated computational model of epiretinal prosthetic vision~\cite{beyeler_axonmap_2019,granley_computational_2021}. Rather than treating electrodes as independent pixels, the model captures characteristic spatial distortions reported by implant users, including elongated phosphene shapes produced by activation of retinal nerve-fiber pathways. We simulated a $10\times10$ epiretinal electrode array, representing a severely resolution-limited prosthetic display. The simulation modeled spatial phosphene appearance only; temporal fading and persistence effects were not included.

A head-tracked virtual camera captured the scene, and each frame was transformed into the corresponding phosphene percept. The spatial model was exported to ONNX and executed on the GPU through Unity Sentis at 90~Hz, allowing the percept to update continuously as participants moved their head. Because the virtual workspace extended beyond the instantaneous field of view of the prosthetic display, targets were often initially invisible and had to be located through active head scanning.

\subsection{Object-Localization Task}

Participants viewed a panoramic virtual workspace containing a semi-circular desk and a set of everyday desk objects that varied in size and visual distinctiveness. Each trial presented a named target together with four distractor objects. Object locations were randomized across trials, and the vertical position of the desk was varied to prevent participants from relying on a fixed scene layout. No visual, auditory, or haptic directional guidance was provided.

At the beginning of each trial, participants were shown the name of a target object. They searched the cluttered SPV-rendered workspace by rotating their head and pulled the trigger on a handheld VR controller held in their preferred hand when they believed the target was centered, ending the trial.
Trials were self-paced: participants had unlimited time to search the scene and confirm their response. The post-trial feedback display was likewise self-paced, remaining visible until the participant pressed the trigger to advance to the next trial.

The primary behavioral outcome was angular localization error, defined as the three-dimensional angular distance between the participant's forward head direction at response and the direction from the participant's head position to the center of the target.
Lower values indicate more accurate localization. 
Completion time was measured from target presentation to trigger response.

\begin{figure}[t]
    \centering
    \includegraphics[width=\linewidth]{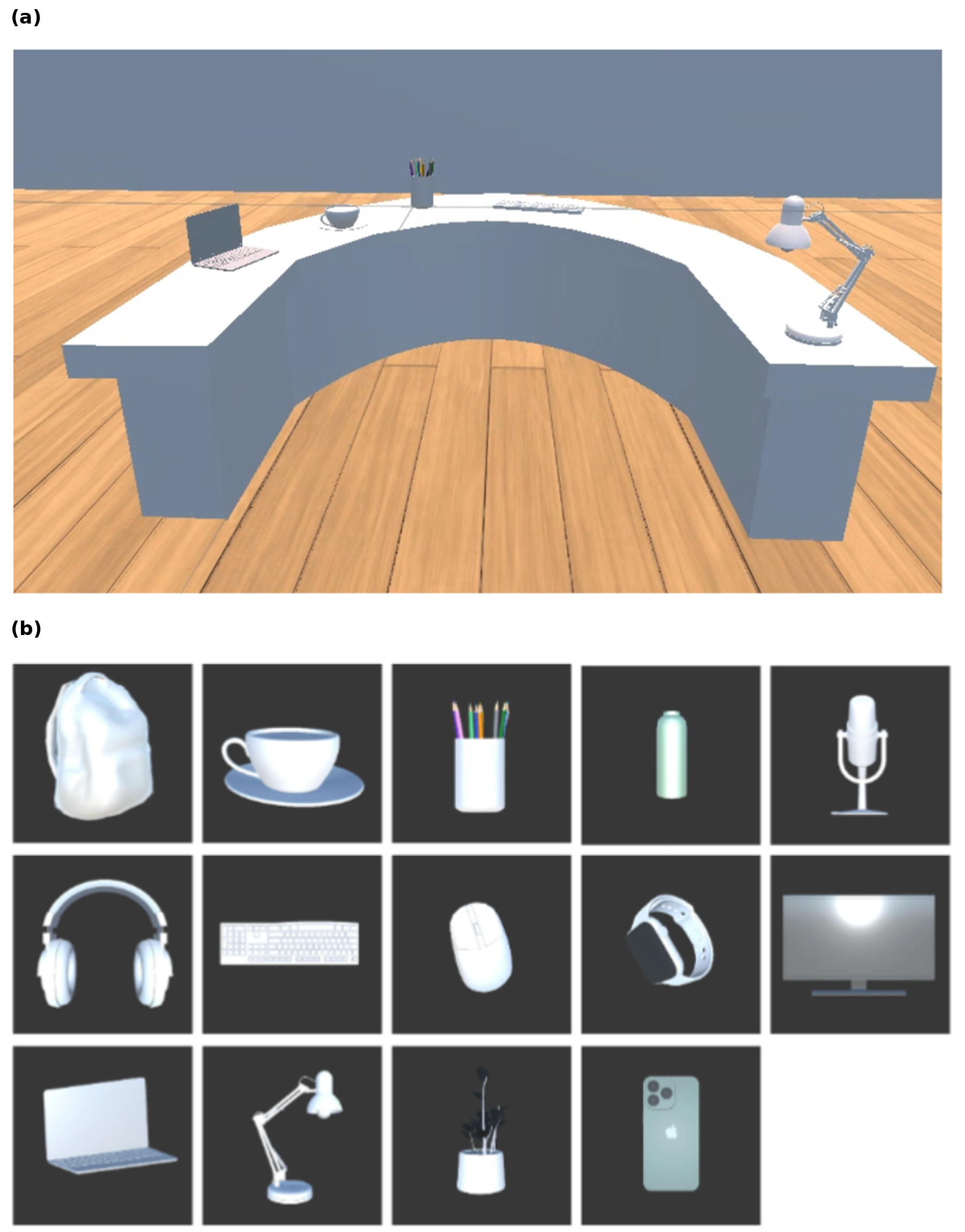}
    \caption{(a) Ground-truth view of the panoramic virtual desk scene before SPV rendering. (b) The 14 everyday objects used as targets and distractors in the virtual search task. On each trial, participants localized a named target among four distractors while viewing the scene through a $10\times10$ simulated phosphene display.}
    \label{fig:objects}
\end{figure}

\subsection{EEG Feedback}

\begin{figure}[t]
  \centering
  \includegraphics[width=\linewidth]{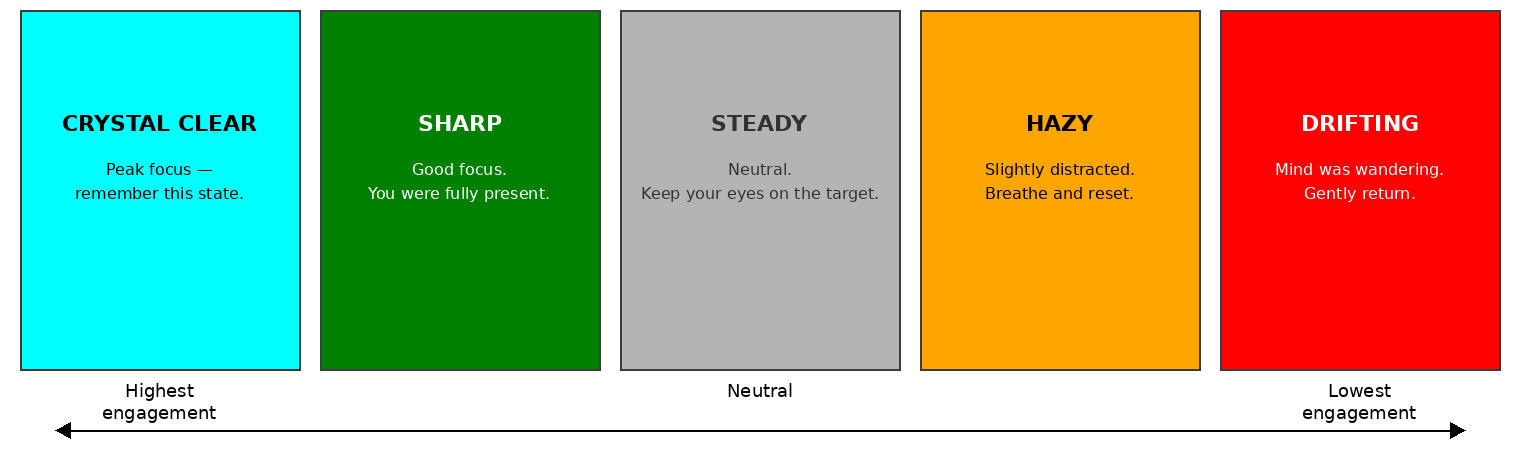}
  \caption{Post-trial neurofeedback display. After each training trial, a full-screen color overlay and descriptive label indicated the participant's engagement level. The neurofeedback group saw values derived from their own EEG; the sham group saw visually identical displays with random values.}
  \label{fig:feedback}
\end{figure}

After each training trial, participants viewed a full-screen color display with a descriptive label indicating their engagement level (Fig.~\ref{fig:feedback}). Five levels ranged from ``Crystal Clear'' (cyan, highest engagement) to ``Drifting'' (red, lowest engagement), each accompanied by a brief coaching message. The display remained visible until the participant advanced to the next trial.

The underlying engagement score was computed as $E = \frac{\beta}{\alpha+\theta}$,
where $\theta$, $\alpha$, and $\beta$ denote power in the 4--8, 8--13, and 13--30~Hz bands, respectively. The mean and standard deviation of the index across the 20 pre-training trials defined a participant-specific baseline. During training, each trial's engagement value was converted to a $z$-score relative to that distribution.

Participants in the neurofeedback group saw feedback derived from their own $z$-scored engagement level. The sham group received visually identical feedback, but its value was sampled uniformly at random from the allowable feedback range rather than derived from the participant's EEG. Feedback appeared only after the search response. This sparse design avoided introducing a competing visual signal into the SPV display, although it also reduced the temporal immediacy of the feedback contingency.

After the session, participants completed a modified NASA Task Load Index assessing mental demand, physical demand, temporal demand, effort, frustration, and perceived performance. 
The performance item was reverse-scored so that higher composite scores indicated greater perceived workload.

\subsection{Statistical Analysis}

The primary intervention analysis compared participant-level pre-to-post changes in angular error between the neurofeedback and sham groups using Welch's two-sample $t$-test; completion time was analyzed analogously. Welch's test was used because group sizes and variances were not assumed to be equal. Phase-specific group means and Cohen's $d$ are reported to characterize the observed differences. Workload and mean training engagement were compared between groups using Welch's tests, with Mann--Whitney $U$ tests as non-parametric sensitivity checks. Given the small feasibility sample, these analyses should not be interpreted as demonstrating equivalence or excluding modest intervention effects. All tests were two-tailed with $\alpha=.05$.

Exploratory trial-level analyses examined whether the engagement index covaried with localization performance; they were not treated as tests of neurofeedback efficacy. Linear mixed-effects models included trial-level engagement, feedback group, and their interaction as fixed effects, with participant as a random effect. Participant-level Spearman correlations between mean training engagement and mean angular error were also computed separately by group. Finally, we descriptively examined heterogeneity within the neurofeedback group using the original responder criteria: a positive slope in relative beta power across training blocks, mean training beta exceeding the participant's pre-training mean by at least one standard deviation, and behavioral improvement from pre- to post-training. Because these criteria were selected post hoc and included behavioral improvement, the responder analysis describes observed trajectories but cannot independently test intervention efficacy.

\section{Results}

\subsection{Behavioral Performance}

\begin{figure*}[t]
  \centering
  \includegraphics[width=\linewidth]{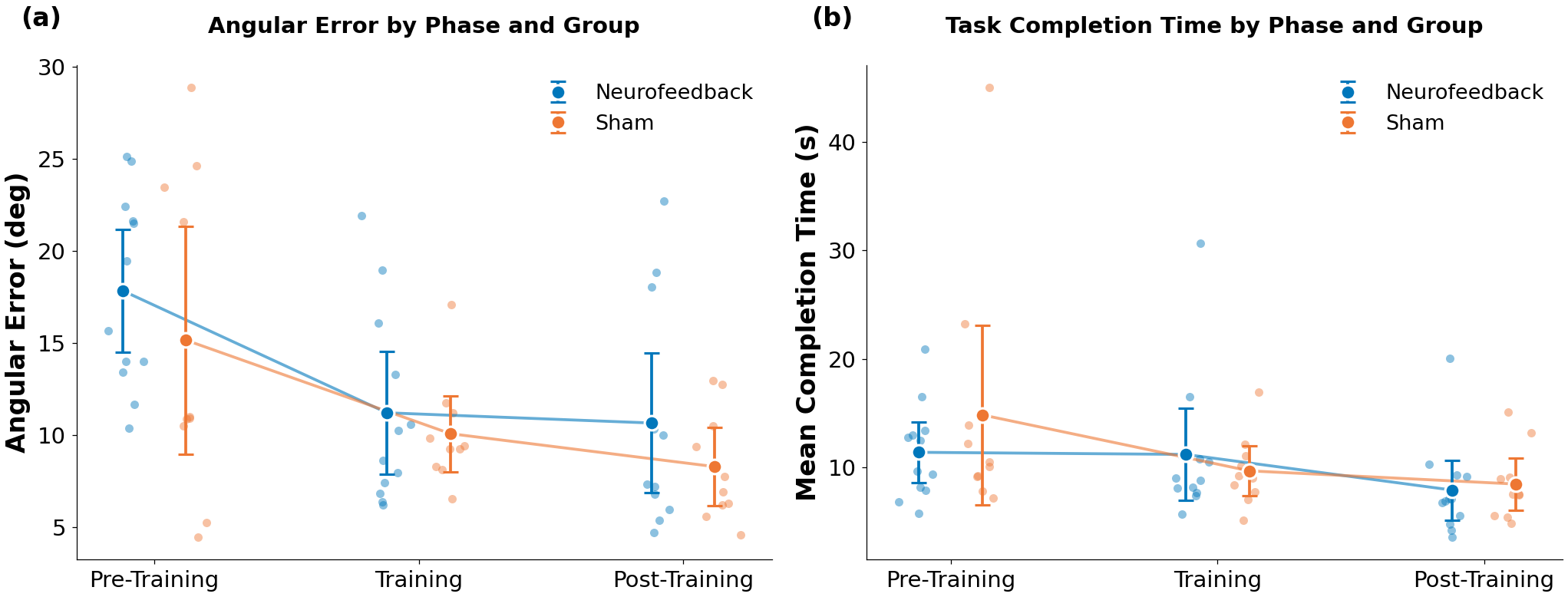}
  \caption{Behavioral performance by phase and feedback condition. Dots denote participants; error bars show 95\% confidence intervals. Mean localization error and completion time decreased across the session, but pre-to-post change did not differ reliably between neurofeedback and sham feedback.}
  \label{fig:behavior}
\end{figure*}

Angular error showed no reliable group difference at pre-training (neurofeedback: $17.84^\circ \pm 5.23^\circ$; sham: $15.15^\circ \pm 8.66^\circ$; $t=0.86$, $p=.405$, $d=0.38$), training (neurofeedback: $11.21^\circ \pm 5.25^\circ$; sham: $10.08^\circ \pm 2.88^\circ$; $t=0.64$, $p=.530$, $d=0.26$), or post-training (neurofeedback: $10.66^\circ \pm 5.95^\circ$; sham: $8.29^\circ \pm 2.97^\circ$; $t=1.21$, $p=.243$, $d=0.49$; Fig.~\ref{fig:behavior}). 

Mean angular error decreased from pre- to post-training in both groups, by $7.18^\circ \pm 5.98^\circ$ in the neurofeedback group and $6.86^\circ \pm 10.13^\circ$ in the sham group. Because the magnitude of improvement was comparable between conditions, the result is consistent with general within-session practice rather than a specific effect of EEG-contingent feedback. Pre-to-post change did not differ between groups ($t=-0.09$, $p=.932$, $d=-0.04$).

Completion time followed the same pattern. There were no reliable group differences at pre-training ($p=.392$), training ($p=.498$), or post-training ($p=.740$), and pre-to-post change did not differ between groups ($t=0.76$, $p=.462$, $d=0.34$). 

Subjective workload also did not differ between groups. NASA-TLX composite scores were similar for neurofeedback and sham (neurofeedback: $2.90 \pm 0.56$; sham: $2.82 \pm 0.50$; $t=0.38$, $p=.707$, $d=0.16$). None of the six workload subscales differed significantly between groups. Effort was the highest-rated subscale in both groups, consistent with the active search demands imposed by the degraded visual display.

\subsection{Engagement Across Training Blocks}

\begin{figure}[t]
  \centering
  \includegraphics[width=\linewidth]{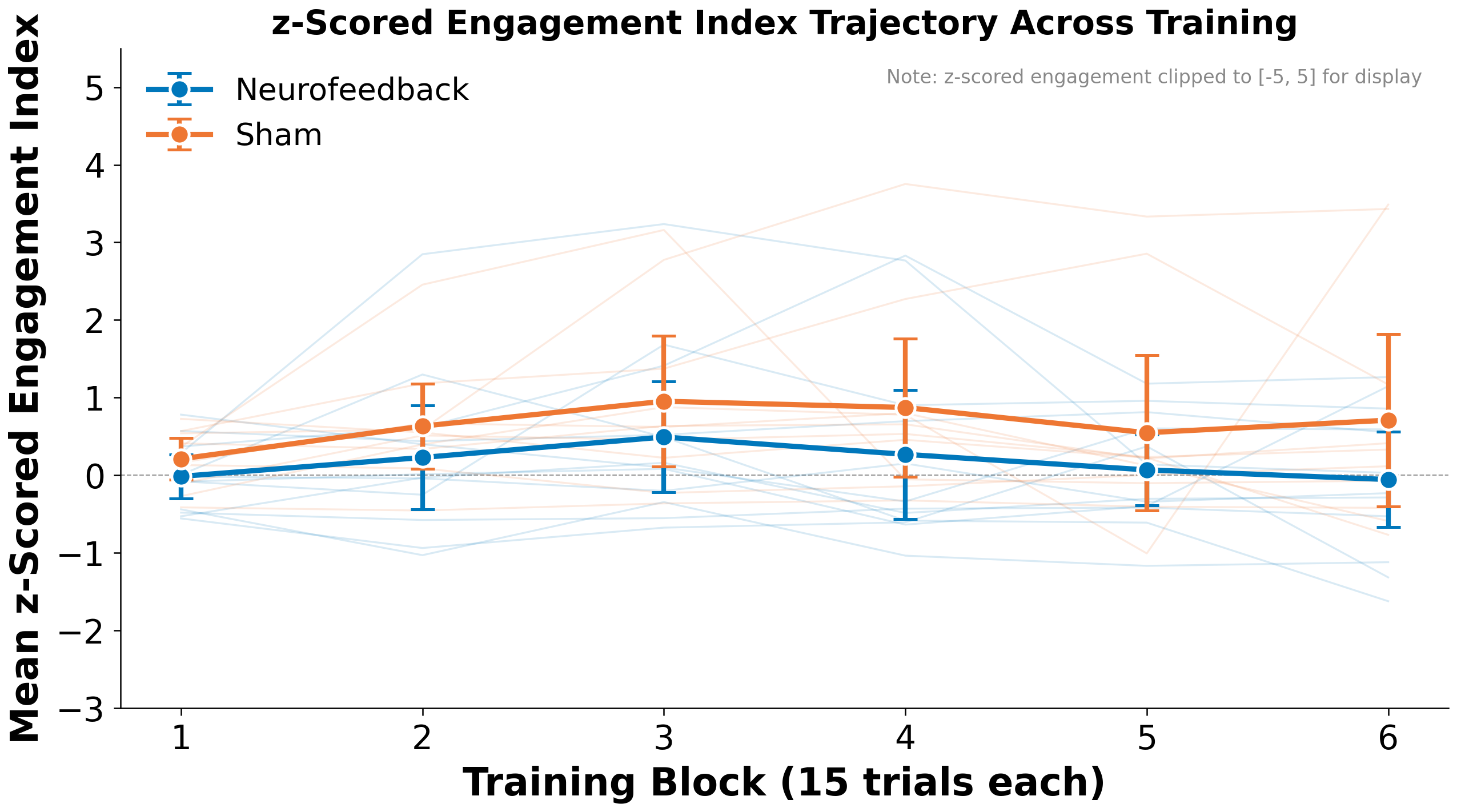}
  \caption{Z-scored engagement index trajectory across six training blocks 
  (15 trials each). Individual participant trajectories are shown as 
  thin lines; group means and 95\% confidence intervals are overlaid. 
  The neurofeedback group did not show systematic upregulation of the engagement 
  index relative to Sham.}
  \label{fig:eeg}
\end{figure}

The raw engagement index averaged across training did not differ between groups (neurofeedback: $0.70 \pm 0.29$; sham: $0.79 \pm 0.43$; $t=-0.53$, $p=.603$, $d=-0.24$). The z-scored engagement index showed no systematic increase across the six training blocks in the neurofeedback group and no reliable divergence from sham (Fig.~\ref{fig:eeg}). Relative beta power showed the same qualitative pattern. A full-session analysis spanning pre-training, the six training blocks, and post-training similarly showed no group-level separation in either the engagement index or relative beta power.
Thus, the trial-level analysis did not provide evidence that the relationship between engagement index and localization error differed between the feedback conditions.

\subsection{Association Between Engagement and Localization Performance}

\begin{figure*}[t]
  \centering
  \includegraphics[width=\linewidth]{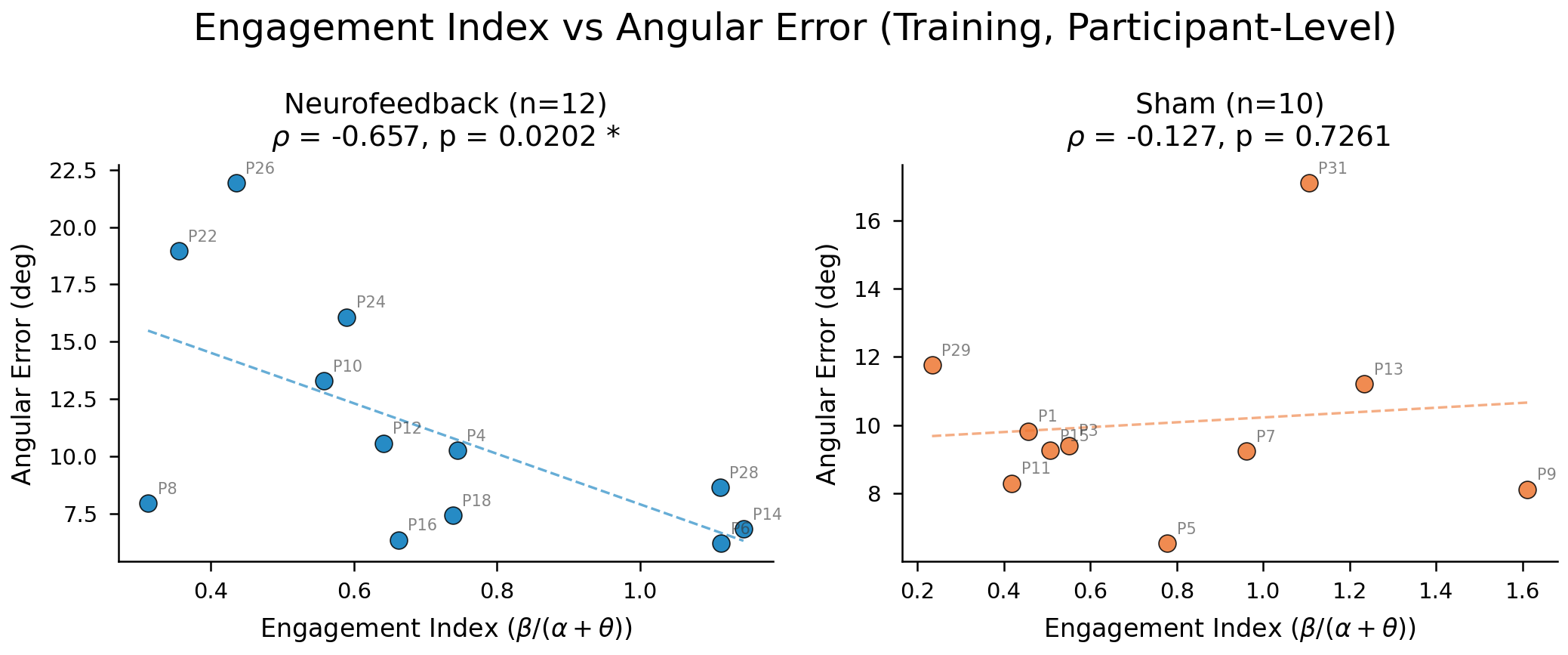}
  \caption{Participant-level relationship between mean engagement index and mean angular error during training. Higher engagement was associated with lower error in the neurofeedback group ($\rho=-.657$, $p=.020$, $n=12$), but not in the sham group ($\rho=-.13$, $p=.726$, $n=10$). The analysis was exploratory, and the group-by-engagement-index interaction in the trial-level mixed model was not significant.}
  \label{fig:individual}
\end{figure*}

We next examined whether the EEG engagement index covaried with localization performance. In the joint trial-level mixed-effects model, the group-by-engagement-index interaction was not reliable ($\beta=-0.22$, $z=-1.06$, $p=.289$). A separate model fit to the neurofeedback group produced convergence warnings, so its standard errors were not interpreted.

At the participant level, higher mean engagement-index values were associated with lower mean angular error in the neurofeedback group, whereas no association was observed in the sham group. This contrast is descriptive: a significant correlation in one group and a nonsignificant correlation in another does not establish that the correlations differ. Given the small group sizes, post-hoc nature of the analysis, and nonsignificant group-by-engagement interaction, the result does not show that contingent feedback altered engagement or caused improved performance.

\subsection{Exploratory Individual Differences in Neurofeedback Response}

Individual beta-power trajectories varied substantially during training. Some participants showed increasing beta power across blocks, whereas others showed stable or declining trajectories. Behavioral improvement was observed across these different patterns, and the small sample did not support identifying a distinct responder subgroup. These descriptive observations motivate prospective studies using pre-specified neural-response criteria, but provide no evidence of responder-specific treatment efficacy.

\section{Discussion}

\subsection{Closed-Loop Neuroadaptive Training Under Simulated Prosthetic Vision}

This study integrated head-tracked SPV, dry-electrode EEG, online computation of an engagement index, and sham-controlled post-trial feedback within a single closed-loop system. Both groups became more accurate over the session, but EEG-contingent feedback did not improve localization accuracy, completion time, subjective workload, or the targeted engagement index relative to sham. The shared improvement most likely reflects repeated exposure, increasing familiarity with the phosphene display and object set, or refinement of visual-search strategies. The study therefore demonstrates the feasibility of the integrated platform, but not a behavioral benefit of the present feedback protocol.

\subsection{Designing Feedback for a Demanding Visual Task}

The absence of a group-level effect may reflect the targeted EEG measure, the short training duration, the feedback design, or the absence of a meaningful intervention effect. The present data cannot distinguish among these explanations. Neurofeedback learning often develops over multiple sessions~\cite{enriquez_geppert2017,sitaram2017}, whereas participants may have learned the localization task relatively quickly.

Presenting feedback between trials prevented it from competing with the degraded visual percept, but weakened the temporal link between cognitive strategy and feedback. Future work could strengthen this contingency through multi-session training, explicit strategy coaching, adaptive targets, or non-visual feedback.

\subsection{Engagement as a Personalization Signal}

The exploratory analyses do not establish the engagement index as either a valid marker of SPV task state or an effective neurofeedback target. The participant-level association in the neurofeedback group was not supported by a significant group-by-engagement interaction and could reflect stable individual differences, behavioral strategy, movement, or other uncontrolled factors. It therefore motivates prospective validation, but not a feedback-specific interpretation.

Signal validity and signal controllability must be established separately. Future studies should first test whether candidate EEG features predict SPV performance after accounting for movement and task duration, and then determine whether those features can be intentionally regulated. Individual variation in beta trajectories also suggests that a single target and feedback mapping may not suit every user.

\subsection{Limitations and Future Work}

The sample ($n=22$) was sufficient only to detect large between-group effects, so the null result does not exclude more modest benefits of EEG-contingent feedback. The study also used sighted participants under simulated prosthetic vision; this supports controlled early-stage testing but does not capture blindness-related search strategies, long-term implant adaptation, or the perceptual variability of electrical stimulation.

The EEG measure has important limitations. Eight-channel dry-electrode recordings during active head movement are vulnerable to motion, muscle activity, and changes in electrode contact, particularly in the beta band. The Galea interface did not expose the underlying preprocessing pipeline, and averaging across electrodes and self-paced trials further reduced spatial and temporal specificity. The engagement index should therefore be interpreted as an online feedback signal rather than a validated neural measure of engagement.

We also did not systematically record the strategies participants used to influence the feedback. Future studies should first validate candidate EEG features against SPV performance, then test whether individualized, multi-session feedback can reliably alter those features and improve behavior. Adaptive difficulty, personalized thresholds, explicit strategy coaching, and non-visual feedback are promising directions.

Overall, the study establishes the feasibility of integrating EEG-contingent feedback with immersive SPV in a sham-controlled protocol. It does not establish the engagement index as a biomarker of SPV performance or demonstrate a behavioral benefit of the present single-session intervention.

\bibliographystyle{abbrv-doi-narrow}

\bibliography{references}
\end{document}